\documentclass[a4paper, amsfonts, amssymb, amsmath, prd, showkeys, nofootinbib, twocolumn, superscriptaddress, colorlinks, linkcolor=refcol, citecolor=refcol, urlcolor=refcol]{revtex4-2}
\usepackage[english]{babel}
\usepackage[utf8]{inputenc}
\usepackage{orcidlink}    
\usepackage[colorinlistoftodos, color=green!40, prependcaption]{todonotes}
\usepackage{bbold}
\usepackage{physics}
\usepackage{makecell}
\definecolor{refcol}{RGB}{178,34,34}
\usepackage{microtype}
\usepackage[T1]{fontenc}
\usepackage{lmodern}
\usepackage{microtype}
\usepackage{booktabs}
\renewcommand{\tr}{\operatorname{Tr}}
\newcommand{\BR}{\mathcal B}
\newcommand{\PG}{P}
\graphicspath{{./figures/}}
\definecolor{red}{rgb}{1,0,0}
 
\definecolor{internationalkleinblue}{rgb}{0.0, 0.18, 0.65}
 
\allowdisplaybreaks

\begin{document}
\title{Glueballs and fractional anomalous determinants at nonzero $\theta$,\\and the decays of the X(2370)}

\author{Francesco Giacosa \orcidlink{0000-0002-7290-9366}}
\email{fgiacosa@ujk.edu.pl}

    \affiliation{Institute of Physics, Jan Kochanowski University, ulica Uniwersytecka 7, P-25-406 Kielce, Poland}
    \affiliation{Institute for Theoretical Physics, Goethe-University,
Max-von-Laue-Straße 1, D-60438 Frankfurt am Main, Germany}

\author{Shahriyar Jafarzade \orcidlink{0000-0002-4353-0760}}
\email{shah.jzade@pitt.edu}
\affiliation{University of Pittsburgh, Pittsburgh, PA 15260, USA}

\author{Győző Kovács \orcidlink{0000-0003-0570-3621}}
\email{kovacs.gyozo@wigner.hun-ren.hu}
    \affiliation{Institute for Particle and Nuclear Physics,
    HUN-REN Wigner Research Centre for Physics, 1121 Budapest, Konkoly–Thege Miklós út 29-33, Hungary}
    \affiliation{Institute of Theoretical Physics, University of Wroclaw, PL-50204 Wrocław, Poland}

\author{Péter~Kovács~\orcidlink{0000-0003-3735-7620}}
\email{kovacs.peter@wigner.hun-ren.hu}
    \affiliation{Institute for Particle and Nuclear Physics, HUN-REN Wigner Research Centre for Physics, 1121 Budapest, Konkoly–Thege Miklós út 29-33, Hungary}

\author{Robert D. Pisarski  \orcidlink{0000-0002-7862-4759}}
\email{pisarski@bnl.gov}
    \affiliation{Department of Physics, Brookhaven National Laboratory, Upton, NY 11973}

\author{Fabian Rennecke \orcidlink{0000-0003-1448-677X}}
\email{fabian.rennecke@theo.physik.uni-giessen.de}
    \affiliation{Institute for Theoretical Physics, Justus Liebig University Giessen, Heinrich-Buff-Ring 16, 35392 Giessen, Germany}
    \affiliation{Helmholtz Research Academy Hesse for FAIR (HFHF), Campus Giessen, Giessen, Germany}


\begin{abstract}
Using fractional anomalous determinants,
we construct a model of the trace and axial anomalies of a $SU(3)$ gauge
theory at nonzero $\theta$ angle through the couplings
to scalar and pseudoscalar glueballs. 
In the pure gauge theory,
we reproduce the $\theta$ dependence of the vacuum energy and the scalar-glueball
mass from the lattice, and estimate the jump in the
topological charge density for the first-order transition
at $\theta = \pi$.  In QCD, the coupling of a pseudoscalar
glueball to fractional anomalous determinants describes all of the four three-pseudoscalar decay
channels of 
the $0^{-+}$  glueball candidate $X(2370)$ measured by the BESIII collaboration.  
Measuring the decay rates for
$\eta \eta \eta$ and $\eta \eta \eta'$ can provide a
stringent test of the model.
\end{abstract}


\hypersetup{
pdftitle={Glueballs and fractional anomalous determinants at nonzero theta, and the decays of the X(2370)},
pdfauthor={F. Giacosa, S. Jafarzade, Gy. Kovacs, P. Kovacs, R. D. Pisarski, F. Rennecke}
}

\maketitle

Quantum Chromodynamics (QCD) possesses two notable
quantum anomalies, which are 
symmetries of the classical Lagrangian that are broken by
quantum fluctuations.
The trace anomaly breaks the classical scale invariance of the Yang--Mills
(YM) sector through dimensional transmutation, generating the nonperturbative
scale $\Lambda_{\mathrm{YM}}$ \cite{Luscher:1993gh}, a negative vacuum energy, and a spectrum of
gluonic bound states, or glueballs \cite{Schechter:1971efy}. The lightest of these is the scalar
glueball \cite{Chen:2005mg}, which provides a natural low-energy realization of the dilatation
degree of freedom \cite{Migdal:1982jp,Gomm:1985ut}. 
The second anomaly is for the axial $U(1)_{A}$ symmetry of the quarks.
This is tied to the topological structure of the $\theta$-vacuum \cite{Bonanno:2025wcv}, and naturally couples quarks to the
channel for the pseudoscalar glueball. 

The effective theory for chiral symmetry breaking
is constructed from the field $\Phi$, where if $q$ are the quark fields, 
$\Phi \sim \overline{q}_L q_R$ is a color singlet.
For $N_f$ massless flavors the flavor symmetry is 
$U(1)_A \times G_{\rm qu}$, with
$G_{\rm qu} =SU(N_f)_L \times SU(N_f)_R$ an exact global symmetry.
Under chiral rotations, $\Phi \rightarrow {\rm e}^{i \alpha_A} \Omega_L \, \Phi \, \Omega_R$,
where $\alpha_A$, $\Omega_L$, and $\Omega_R$ are elements of
$U(1)_A \times G_{\rm qu}$.
In QCD the chiral symmetry is approximate,
and $N_f = 2+1$, as the strange quark is much heavier
than the up and down quarks.

Any potential for the chiral effective Lagrangian involves terms such as 
powers of traces of $\Phi^\dagger \Phi$, such as $\Tr \Phi^\dagger \Phi$,
$\Tr (\Phi^\dagger \Phi)^2$, {\it etc.}.  But these
are all invariant
under $G_{\rm qu}$ {\it and}
the anomalous $U(1)_A$ symmetry.  The unique operator
which is invariant under $G_{\rm qu}$, but {\it not} $U(1)_A$, is
the anomalous determinant of Kobayashi, Maskawa, and 't Hooft
\cite{Kobayashi:1970ji,Kobayashi:1971qz,Kobayashi:1973fr,Maskawa:1974,tHooft:1976rip,tHooft:1976snw}, ${\rm e}^{i \theta} \det \Phi$.
Integral powers of the anomalous determinant enter into the 
chiral effective Lagrangian as
\begin{equation}
\left({\rm e}^{i\theta}\det\Phi\right)^{{\cal Q}}
\, + \, \left({\rm e}^{- i\theta}\det\Phi^\dagger\right)^{{\cal Q}} \; ; 
\label{eq:anomalous}
\end{equation}
${\cal Q}$ is the instanton number, and in the dilute gas approximation, is an integer.
Such operators are essential in describing 
the breaking of the axial $U(1)_A$ symmetry.  In vacuum, it is clear
that for massless quarks, the $\theta$-angle can be eliminated by a $U(1)_A$
rotation.  In vacuum, operators with ${\cal Q} \neq 0$
split the mass of the iso-singlet
$\eta^{\prime}(958)$ from the octet of $\pi$, $K$, and $\eta$ mesons
\cite{Kobayashi:1970ji,Kobayashi:1971qz,Kobayashi:1973fr,Maskawa:1974,tHooft:1976rip,tHooft:1976snw,Rosenzweig:1981cu,tHooft:1986ooh,Bass:2018xmz,Pisarski:2019upw,Rennecke:2020zgb,Giacosa:2024epf,Giacosa:2026lal} and 
is essential at nonzero temperature and/or density \cite{Pisarski:2024esv,Pisarski:2024esv,Giacosa:2024orp}. 
A generalized polydeterminant describes
the splitting of heterochiral mesons with spin $J \geq 1$
\cite{Giacosa:2017pos,Giacosa:2023fdz,Giacosa:2025dvu}. 

It is expected that QCD with three colors, $N_c = 3$, is qualitatively
similar to large $N_c$ \cite{tHooft:1973alw,Witten:1979kh,Giacosa:2024scx}.
Holding $N_f$ fixed as $N_c \rightarrow \infty$, the only breaking of the
axial $U(1)_A$ is a mass for the $\eta'$, which is light, $m_{\eta'}^2 \sim N_f/N_c$.
This is represented by the Witten-Veneziano term, 
$\sim N_f/N_c |\log \det U|^2$, where $U$ is the usual chiral field for pions
and the $\eta'$
\cite{Witten:1979vv,Witten:1980sp,Veneziano:1979ec,DiVecchia:1980yfw}
\footnote{
$U$ is constructed from $\Phi \equiv(f_{\pi}/2)U$, where $U = \Phi/\sqrt{\Phi^\dagger \Phi}$, and $U^\dagger U = {\bf 1}$.
}.
Further, if $\det \Phi$ dominates near the phase transition for the restoration
of the chiral symmetry, it predicts a strongly first order chiral phase transition
for three light flavors \cite{Pisarski:1983ms}.  There is no evidence
for such a first order transition from numerical simulations on the lattice
\cite{Cuteri:2021ikv,Pisarski:2024esv,Ding:2026blh};
the chiral transition for $N_f=3$ may even be "beyond Landau" \cite{Zhang:2026zpa,Pisarski:2026}.

In this Letter we show how
the ordinary determinant emerges as one element of a general tower of
{\it fractional} powers of the anomalous determinant from the $\theta$-dependence of the pure glue theory \cite{Witten:1980sp}.  This is identical
to Eq.\ \ref{eq:anomalous}, except that now ${\cal Q}$ is fractional,
\begin{equation}
{\cal Q} = \frac{n}{N_c} \; , \; n = \pm 1, \pm 2 \ldots
\label{frac_top_chg}
\end{equation}
This is motivated by Ref.\ \cite{Pisarski:2026}, which proposed a generalization
of the Witten-Veneziano term for arbitrary values of $N_f/N_c$.
Here we construct the anomalous interactions of (pseudo)scalar
glueballs with mesons by interactions involving fractional anomalous determinants.
In terms of the nonlinear field $U$,
the fractional anomalous determinants become $\sim (\det U)^{{\cal Q}}$,
as seen on a femto-torus \cite{Hayashi:2024qkm,Pisarski:2026}.

We start with the pure glue theory. 
The Lagrangian is the sum of the usual action, which is invariant
under the discrete symmetry of $CP$, plus a piece proportional to the
$\theta$ angle, which violates $CP$:
\begin{equation}
{\cal L}_\theta = \frac{1}{2}\, \Tr \, G_{\mu\nu}G^{\mu\nu}
- i \, \theta\, {\cal Q}\; ; \;
 {\cal Q}=\frac{g^{2}}{32\pi^{2}}%
\epsilon^{\mu\nu\rho\sigma}
\Tr \, G_{\mu\nu} \, G_{\rho\sigma} \; ; 
\end{equation}
$D_\mu = \partial_\mu - i g [A_\mu, ] $ is the covariant derivative in the adjoint representation, $A_\mu = A_\mu^2 \, t^a$,
with $\Tr\,  t^a t^b = \delta^{a b}/2$,
$G_{\mu\nu}=[D_\mu,D_\nu]/(- i g)$ is the gluon field strength tensor, 
and $g$ the strong coupling constant.

The topological charge density is given by ${\cal Q}$.  This is 
a total divergence, whose integral over spacetime is the instanton number.  
On a compact manifold the instanton number is an integer,
but if the boundary of the manifold is twisted by
elements of the center of the $SU(N_c)$ gauge group, which is $Z(N_c)$,
then ${\cal Q}$ is fractional, as in Eq. \ref{frac_top_chg}.
This is why 
fractional powers of the anomalous determinant appear
in the phase with the spontaneous breaking of chiral symmetry
\cite{Pisarski:2026}. 

Of course in QCD $\theta$ is very small, $\theta < 10^{-9}$ \cite{Baker:2006ts,Abel:2020pzs}.
Nevertheless, it is useful to consider the properties of
non-Abelian gauge theories, without dynamical quarks, at nonzero $\theta$.  
Around $\theta=0$, one can expand the vacuum energy, $E(\theta)$,
and the mass of the lightest excitation, which is a scalar
glueball, in powers of $\theta^2$:
\begin{align}
E_{\mathrm{vac}}(\theta)-E_{\mathrm{vac}}(0)  & =\frac{\chi_{\mathrm{YM}}}%
{2}\theta^{2}\left(  1+b_{2}\theta^{2}+b_{4}\theta^{4}+...\right)\,,
\label{evaclatt} \\
 M_{G}(\theta)  =M_{G}(0)&\left(  1+m_{2}\theta^{2}+m_{4}\theta^{4}%
+...\right) . 
\end{align}
Both quantities scale naturally at large-$N_{c}$: the quadratic
coefficient $\chi_{\mathrm{YM}},$ the
topological susceptibility for the pure glue, Yang-Mills
theory, scales as
$N_{c}^{0}$; $b_{2}$ and $m_{2}$ as $N_{c}^{-2}$; and $b_{4}$ and $m_{4}$ as $N_{c}^{-4},$ {\it etc.}. 
This scaling is confirmed by numerical
simulations on the lattice.  For three colors 
\cite{Bonati:2015sqt, Bonanno:2024ggk},
\begin{align} \label{bglatt}
b_{2}^{lat}  & =-0.0216\pm0.0015\text{ , }\left\vert b_{4}^{lat}\right\vert
\lesssim4\cdot10^{-4}\text{ }
\end{align}
\begin{align} 
m_{2}^{latt}  & =-0.0083\pm0.0023 \text{ .} \label{mglatt}
\end{align}
The gluon Lagrangian $\mathcal{L}$ is dilatation invariant, but 
this is broken by the trace anomaly, as
 $\partial_{\mu}D^{\mu}=- 11N_{c} \alpha_s/(12 \pi)%
\Tr \, G_{\mu\nu} G^{\mu\nu}\neq0,$ with $\alpha_{s}%
=g^{2}/4\pi$ and the dilaton current $D^\mu(x) = x_\nu T^{\mu\nu}(x)$, where $T^{\mu\nu}$ is the energy-momentum tensor.  In the low energy domain, this can be described by a
single dilaton field, subject to the potential \cite{Schechter:1980ak,Migdal:1982jp,Gomm:1985ut}:
\begin{equation}
V_{\mathrm{dil}}(G)=\frac{\lambda_{G}}{4}\, G^{4}
\left(  
\ln\left(
\frac{G}{\Lambda_{G}}\right)-\frac{1}{4}\right)  , \label{eq:dilaton}%
\end{equation}
where $\lambda_{G}\sim N_{c}^{-2}$ is dimensionless and $\Lambda_{G}%
\sim\Lambda_{G}N_{c}$ includes the renormalization
mass scale $\Lambda_G$. Since 
$\partial_{\mu}D_{G}^{\mu}=-\lambda_{G} \, G^{4}/4,$ the correspondence
\begin{equation}
\frac{\lambda_{G}}{4}G^{4}\equiv
\frac{11N_{c}}{12}\frac{\alpha_{s}}{\pi
} \, \Tr \, G_{\mu\nu}G^{\mu\nu} = \frac{11N_c}{24}\mathcal{G}^2
\label{eq:defgc}
\end{equation}
follows. Using this result and the lattice values of Ref.\ \cite{Chen:2005mg}, we determine the parameters of
$V_{\mathrm{dil}}(G)$ as
\begin{align}
m_{G}=M_{G}(0) &=\sqrt{\lambda_{G}}\Lambda_{G} =    1.71\pm0.09\text{ GeV} \nonumber \\
\left\langle 0\left\vert g^{2}
\Tr \, G_{\mu\nu}G^{\mu\nu}\right\vert
0^{++}\right\rangle &=    \frac{48\pi^{2}}{11N_{c}}\lambda_{G}\Lambda_{G}%
^{3}\nonumber = 31.2\pm6.4 \text{ GeV}^{3}\,.
\end{align}
so that
\begin{equation}
\lambda_{G}=21\pm11\text{ , }\Lambda_{G}=0.372\pm0.086\text{ GeV} \text{ .}
\end{equation}
Two consequences are relevant. The vacuum energy
\begin{equation}
B=-E_{\mathrm{vac}}(0)=\frac{\lambda_{G}}{16}\Lambda_{G}^{4}=0.025\pm
0.011\text{ GeV}^{4}%
\end{equation}
fits well with the phenomenological range $B=0.0142$-$0.0263$ GeV$^{4}$
reported in Ref. \cite{Barnafoldi:2007nm}. From Eq. \ref{eq:defgc}, the gluon condensate
\begin{equation}
\left\langle\mathcal{G}\right\rangle_0=\frac{24}{11N_{c}}\frac{\lambda_{G}}{4}\Lambda_{G}^{4}
=0.073\pm0.031\ {\rm GeV}^{4}
\end{equation}
is in very good agreement with the recent lattice
measurement of $0.076\pm0.004$  extracted from  Ref.~\cite{Ayala:2020pxq} (Ref.~\cite{Bali:2014sja} reports a similar value,
$\simeq0.077$ GeV$^{4}$, but with a large error; Ref. \cite{DiGiacomo:1998nu}  
quotes 
$0.14\pm0.02$ GeV$^{4}$). 
The corresponding values with dynamical quarks are smaller
\cite{DElia:1997sdk,DiGiacomo:1998nu}.
Altogether, these are reasonable values
from such an elementary model.

We can go to arbitrary $\theta$ by coupling the dilaton to an even function $f_{G}(\theta/N_c)$ with $f_{G}(0)=0$,
%
\begin{equation}
V(G,\theta)=V_{\mathrm{dil}}(G)+\frac{G^{4}}{2}f_{G}
\left( \frac{\theta}{N_c}\right) \; .
\label{fg}
\end{equation}
The function must be even to ensure CP invariance up to the $\theta$ term and the argument $\theta/N_c$ ensures the correct large-$N_{c}$ scaling \cite{Witten:1980sp}, see also \cite{Gabadadze:2002ff,Vicari:2008jw,Sannino:2026wgx}. We can express $f_G$ as a Fourier series,
\begin{equation}
f_{G}\left( \frac{\theta}{N_c}\right)
= \sum _{n=1}^{\infty}\lambda_{G\theta}^{(n)}\left(  1-\cos\left( \frac{n\theta}{N_c}\right)  \right) \; ,
\label{eq:Gtheta}%
\end{equation}
where $\lambda_{G\theta}^{(n)} \sim N_c^{-2}$.
This is particularly natural because it ensures periodicity of the QCD $\theta$ vacuum, while retaining the expected large-$N_c$ dependence on $\theta/N_c$ \cite{Gabadadze:2000vw,Gabadadze:2002ff}. 

A simplified model is obtained by retaining only the lowest term, 
$f_{G}(\theta/N_c)$
$=\lambda_{G\theta}\left(  1-\cos\left( \theta/N_c \right)  \right)  $. It is
interesting to check its numerical consequences. Minimization gives
\begin{equation}
G_{0}(\theta)=\Lambda_{G}
\exp\left(-\frac{2\lambda_{G\theta}}{\lambda_{G}}
\left(  1-\cos \frac{\theta}{N_c}\right)\right)\text{ ,} \label{eq:GthetaVEV}%
\end{equation}
so that the $\theta$ vacuum induces a backreaction on the gluon condensate.
Consequently,
\begin{equation}
E_{\mathrm{vac}}(\theta)=-B\,
\exp 
\left(
-\frac{8\lambda_{G\theta}}{\lambda_{G}}
\left(  
1-\cos \frac{\theta}{N_c}
\right)
\right)\; .
\label{eq:Evactheta}%
\end{equation}
Expanding, the dominant term of $E_{vac}(\theta)-E_{vac}(0)$ is proportional to $\cos \theta/N_c$, as postulated in Refs. \cite{Gabadadze:2000vw,Gabadadze:2002ff}.
Fixing $\lambda_{G\theta}$ from the YM topological susceptibility leads to
\begin{equation}
\chi_{\mathrm{YM}}=\frac{\lambda_{G\theta}\Lambda_{G}^{4}}{2N_{c}^{2}},
\end{equation}
which is independent of $N_{c}$. This is due to the explicit dependence via
$\theta/N_{c}.$ Without it, the wrong result $\chi_{\mathrm{YM}}\sim
N_{c}^{2}$ would follow. For $N_{c}=3$ lattice finds $\chi_{\mathrm{YM}}
^{1/4}=198.1(0.7)(2.7)\ \mathrm{MeV}$  \cite{Durr:2025qtq}, hence $\lambda_{G\theta}=1.4\pm1.3$.
Expanding: 
\begin{align}
b_{2}  &  =-\frac{1}{12N_{c}^{2}}-\frac{\chi_{\mathrm{YM}}}{4B}\overset
{N_{c}=3}{=}-0.0245(65),\label{eq:b2m2}\\
\qquad m_{2}  &  =-\frac{\chi_{\mathrm{YM}}}{8B}\overset{N_{c}=3}%
{=}-0.0076(33),
\end{align}
thus $b_2-2m_2=-1/(12N_c^2)$.  The values are in excellent agreement with lattice, see Eqs. (\ref{bglatt}) and (\ref{mglatt}). Interestingly, the scaling $b_{2} \sim N_c^{-2}$ is confirmed by lattice simulations \cite{Bonati:2016tvi}. 
The model also 
predicts
\begin{align*}
b_{4}  &  =\frac{1}{360N_{c}^{4}}+\frac{\chi_{\mathrm{YM}}}{24BN_{c}^{2}%
}+\frac{\chi_{\mathrm{YM}}^{2}}{24B^{2}}\overset{N_{c}=3}{=}(4.7\pm
2.5)\times10^{-4},\\
m_{4}  &  =\frac{\chi_{\mathrm{YM}}}{96BN_{c}^{2}}+\frac{\chi_{\mathrm{YM}%
}^{2}}{128B^{2}}\overset{N_{c}=3}{=}(1.0\pm0.6)\times10^{-4} .
\end{align*}
The former matches the lattice upper limit of Eq. (\ref{bglatt}), while the latter is a prediction.

\begin{figure}[t]
\centering
\includegraphics[width=0.45\textwidth]{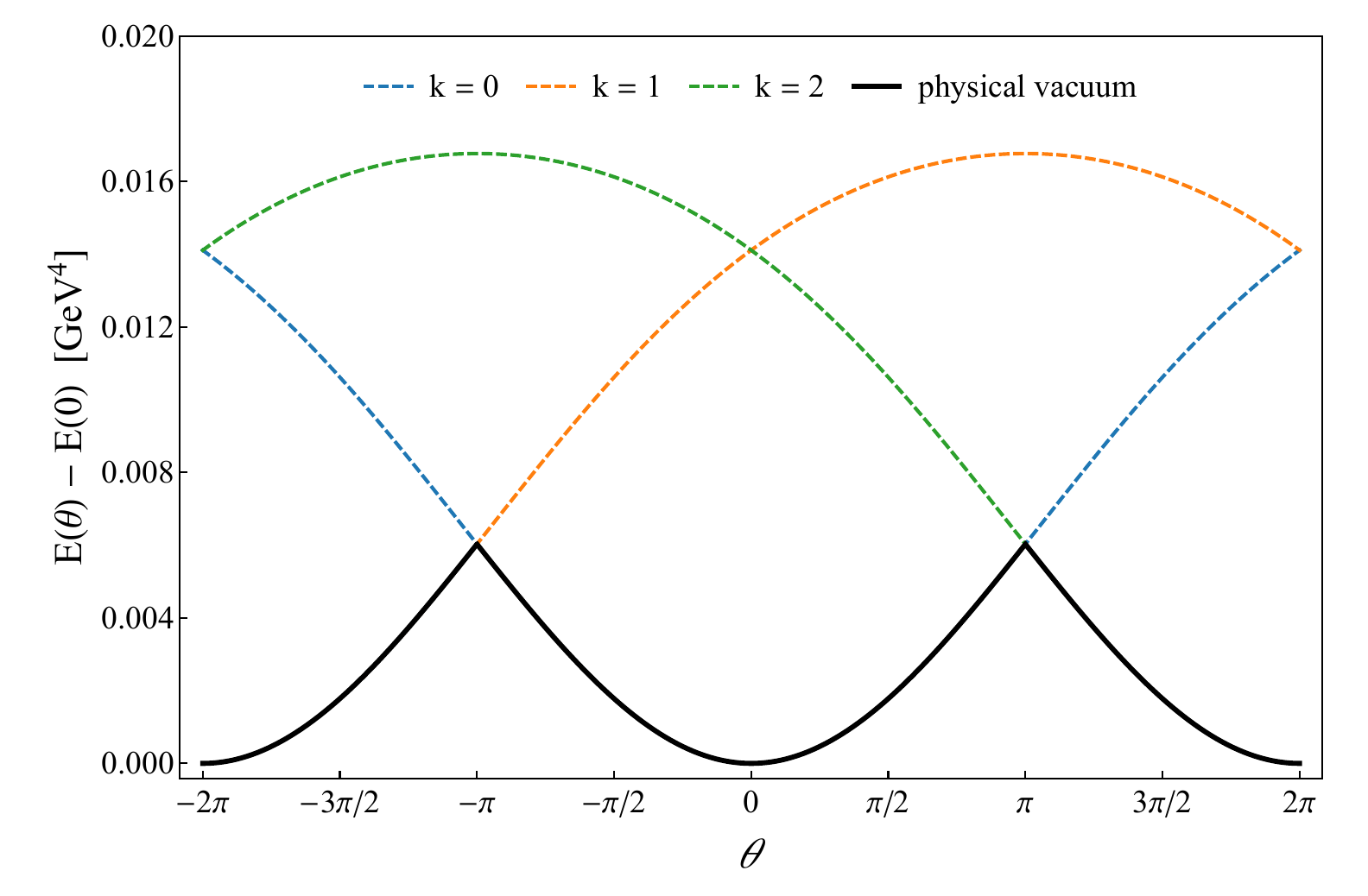}
\caption{
The three branches of the vacuum energy for three colors in the pure
gauge theory, with the true vacuum energy denoted by the solid black line.
Periodicity in $\theta\rightarrow\theta+2\pi$ 
is enforced by jumping branches.  These cross at
$\theta=\pi$, where there is a first-order transition.  
}
\label{fig:YMbranches}
\end{figure}

Away from the
small-$\theta$, as for large $N_c$
\cite{Witten:1979vv,Witten:1980sp,Veneziano:1979ec,DiVecchia:1980yfw},
the periodicity of $\theta$ in $\theta \rightarrow \theta + 2\pi$
is realized through a potential which has multiple branches.
For three colors there are three branches, with
$\theta\rightarrow\theta+2\pi k$, with $k=0,1,2$, in the argument
$(\theta+2\pi k)/N_c$.  The physical vacuum energy is then given by
the lowest branch,
$E_{\rm vac}(\theta)=\min_{k=0,1,2} E_k(\theta)$, see Fig.~\ref{fig:YMbranches}.  While each
individual branch has period $6 \pi$, a shift
$\theta \rightarrow \theta + 2\pi$ simply permutes the branches, so the
physical vacuum energy is periodic in $2\pi$.  

The two branches for $k = 1$ and $2$ meet at $\theta = \pi$.
The energy is continuous at $\theta = \pi$, but the derivative of the energy with
respect to $\theta$ is not. The topological charge density,
${\cal Q} =\partial E_{\rm vac}/\partial\theta$, flips sign at
$\theta = \pi$
\cite{Gabadadze:2000vw,Gabadadze:2000vw}.  
For three colors, in our model we find that at $\theta = \pi$,
\begin{equation}
|\Delta {\cal Q}|
=3\sqrt{3}\,\chi_{\mathrm{YM}}\, 
{\rm e}^{ - 9\chi_{\mathrm{YM}}/2B }
\simeq 6.1\times10^{-3}\ {\rm GeV}^4 .
\end{equation}
For comparison, quadratic branches give
$|\Delta {\cal Q}|=2\pi\chi_{\mathrm{YM} }\simeq9.6\times10^{-3}$ GeV$^4$,
while extrapolating the lattice $b_2$ result to $\theta=\pi$ gives
$\simeq5.5\times10^{-3}$ GeV$^4$. 
A holographic model
yields a similar value, 
$\simeq6.4\times10^{-3} {\rm GeV}^4$
\cite{Bigazzi:2015bna}.

Expanding about $\theta = 0$, 
the gluon condensate reads:
\begin{equation}
\langle \mathcal{G}^{2}\rangle_{\theta}
= \left( 1+c_{2}\theta^{2}+c_{4}\theta^{4}+\ldots\right)
\langle \mathcal{G}^{2}\rangle_{0} \; .
\end{equation}
For three colors, 
\begin{align}
c_{2} &  =-\frac{4\lambda_{G\theta}}{\lambda_{G}N_{c}^{2}}=-\frac
{\chi_{\mathrm{YM}}}{2B}=4m_{2} = \, -0.030\pm0.013,\nonumber \\
c_{4} &  =\frac{\lambda_{G\theta}}{3\lambda_{G}N_{c}^{4}}+\frac{8\lambda
_{G\theta}^{2}}{\lambda_{G}^{2}N_{c}^{4}}=-\frac{m_{2}}{3N_{c}^{2}}+8m_{2}%
^{2}
= \, 7.4\times10^{-4}.
\nonumber
\end{align}
The $\theta$ dependence of the gluon condensate was not evaluated on the
lattice but has been studied in other theoretical approaches, Refs.~\cite{Aitken:2018mbb,Halperin:1997vj}. In the former, the
semiclassical analysis of Yang--Mills theory on $\mathbb{R}^{3}\times S^{1}$
leads to the values $c_{2}%
^{\mathrm{semiclass}}\simeq-0.0556$ and $c_{4}^{\mathrm{semiclass}}%
\simeq5.14\times10^{-4}$, which is close to the values in our model.

The pseudoscalar gluonic degree of freedom can be incorporated through terms
\begin{equation}\label{eq:pot}
V=V_{\rm dil}(G)
+ \frac{G^4}{2}\,  f_G \, 
\left(
\frac{\theta}{N_c}
\right)
+\frac{\lambda_{GP}}{2}G^2P^2
-\frac{G^3P}{3}f_P\!\left(\frac{\theta}{N_c}\right)
\end{equation}
with $f_{P}(\theta/N_c)$ an odd function fulfilling $f_{P}(0)=0$:
\begin{equation}
f_{P}\left( \frac{\theta}{N_c}\right)\text{ }= \sum _{n=1}^{\infty}\lambda_{P\theta}^{(n)}  \sin
\left( 
\frac{n\theta}{N_c}
\right)   \text{ .}
\label{eq:pexp}
\end{equation}
Eqs. \ref{eq:pot} and \ref{eq:pexp} lead to the pseudoscalar glueball coupling to mesons; see below.

With dynamical quarks in QCD, one takes $\theta \rightarrow \theta + \arg\det M$, with the quark mass matrix $M$, in order to account for axial rotations of the quark fields. It is hence natural to introduce (pseudo)scalar mesons to Eq.\ (\ref{eq:pot}) through
\begin{equation}
\theta\;\rightarrow\; \Theta
= 
\theta - i \ln \left(\frac{\det\Phi}{|\det\Phi|}\right)\;.
\label{eq:X}%
\end{equation}
where $\Phi=\mathcal{S}+i\mathcal{P}$ contains the $N_f \times N_f$ matrices with scalar and pseudoscalar mesons. As before, different branches are obtained with
$\Theta\rightarrow\Theta+2\pi n$.  
The vacuum is fixed by the pion and kaon decay constants:
$ \mathcal{S}_0=\Phi_0=\frac12\operatorname{diag}
\left(f_\pi,f_\pi,2f_K-f_\pi\right)$,
with $f_\pi=92.2$ MeV and $f_K=110$ MeV ($\Phi_0=\frac{f_{\pi}}{2} \mathbf{1}$ in the flavor-symmetric limit); see supplemental material for further technical details. 

After the shift, Eq. (\ref{fg}) becomes
%
\begin{align}
\nonumber f_{G}\left(\frac{\Theta}{N_c}\right)
&=\sum_{n=1}^{\infty}\lambda_{G\theta}^{(n)}
\left[  1-\cos
\left(
\frac{n \Theta}{N_c} 
\right)
\right]\\
&= \sum_{n=1}^{\infty}\lambda_{G\theta}^{(n)}
\left[
 e^{i n\theta/N_c}
 \left(\frac{\det\Phi}{|\det\Phi|}\right)^{n/N_c}
 +\mathrm{h.c.}
\right]
\end{align}

The potential $V(G,\Theta)$ reduces to the Witten-Veneziano
mass relation in the chiral limit,
$m_{\eta_0}^2=2N_f\chi_{\mathrm{YM}}/f_{\pi}^2\sim N_c^{-1}$
\cite{Witten:1979vv,Veneziano:1979ec}.

Interestingly, the quadratic  gluonic coupling 
\begin{equation}
f_G\ \propto \left(\frac{\Theta}{N_c}\right)^2
=4\sum_{n=1}^{\infty}\frac{(-1)^{n+1}}{n^2}
\left[1-\cos\left(\frac{n\Theta}{N_c}\right)\right] 
\end{equation}
shows that the quadratic Witten form can also be represented by the
same fractional-determinant phase harmonics, with fixed Fourier coefficients. This term can be related to a generalized
``quenched'' topological susceptibility \cite{DiVecchia:2026mrz}, where the
matching must account for screening by the explicitly retained
gluonic modes, so that the susceptibility for the pure gauge
theory reproduces
$\chi_{\mathrm{YM}}$. 

After the shifts $G\rightarrow G_0+G$ and Eq. \ref{eq:X}, the scalar-glueball interaction
Lagrangian contains, up to terms involving only gluonic fields
\begin{equation}
G_0^3G\sum_{n=1}^{\infty}\lambda_{G\theta}^{(n)}
\left[
 e^{in \theta/N_c}
 \left(\frac{\det\Phi}{|\det\Phi|}\right)^{n/N_c}
 +\mathrm{h.c.}
\right].
\label{eq:Gmesons}
\end{equation}
Similarly, at $G=G_0$, the pseudoscalar-glueball interaction reads
\begin{equation}
iP\sum_{n=1}^{\infty}b_n
\left[
 e^{in \theta/N_c}
 \left(\frac{\det\Phi}{|\det\Phi|}\right)^{n/N_c}
 -\mathrm{h.c.}
\right],
\label{eq:Pmesons}
\end{equation}
where $b_n=-G_0^3\lambda_{P\theta}^{(n)}/6$ with the sign convention
of the potential above. For $n=N_c$, replacing the determinant
modulus by its vacuum value recovers, up to an overall coupling,
$iP[\det\Phi- \det \Phi^\dagger]$, studied in Ref.~\cite{Eshraim:2012jv}.
The normalized interaction, however, must retain the field dependence
of $|\det\Phi|$ when its decay vertices are calculated.

On the vacuum branch at $\theta=0$, the direct three-pseudoscalar
vertices follow by setting
$\Phi=\Phi_0+i\mathcal{P}$ and expanding the complete normalized
expression. The terms cubic in $\mathcal{P}$ are
\begin{align}
\operatorname{Im}
\left(\frac{\det\Phi}{|\det\Phi|}\right)^{n/N_c}
\!\!=&-\tfrac{n}{3N_c}
\Tr\left[(\Phi_0^{-1}\mathcal{P})^3\right] \nonumber \\
&-\tfrac{n^3}{6N_c^3}
\left[\Tr(\Phi_0^{-1}\mathcal{P})\right]^3 + \ldots
\end{align}
Remarkably, the expansion of the inverse determinant modulus cancels
the $\Tr [(\Phi_0^{-1}\mathcal{P})^2] \Tr  (\Phi_0^{-1}\mathcal{P})$) term from the numerator. Hence only the single-trace and triple-trace structures displayed
above survive at cubic order. 
These interactions can be tested using the pseudoscalar-glueball
candidate $P \equiv X(2370)$
with a mass of $M_X=2.359$ GeV  \cite{BESIII:2023wfi,BESIII:2026rzt,BESIII:2026mvn}, in line with lattice simulations \cite{Cherman:2016hcd}.

since $\Tr(\Phi_0^{-1}\mathcal{P})$ contains only $\eta_N$ and
$\eta_S$, its cube does not contribute to
$K\bar K\pi$, $\eta\pi\pi$, $\eta'\pi\pi$, or $\eta'K\bar K$.
Their direct decay amplitudes arise entirely from
$\Tr(\Phi_0^{-1}\mathcal{P})^3$, with the common coefficient
$\sum_n n b_n$. All individual harmonics, and any combination with
nonzero common coefficient, therefore give the same ratios.

\begin{center}
\begin{tabular}{c|cc}
channel & exp. & theory\\
\hline
$\pi \, K \, \overline{K}$ & $3.25^{+0.77}_{-0.79}$ & $2.90$\\
$\pi\, \pi\, \eta$ & $3.20^{+0.91}_{-1.00}$ & $3.89$\\
$ \pi\, \pi\, \eta'$ & $1.94^{+0.33}_{-0.88}$ & $1.99$\\
$K \, \overline{K} \, \eta' $ & $0.39\pm0.11$ & $0.278$\\
\hline
$\chi^2/\mathrm{d.o.f.}$ & -- & $1.82/3$
\end{tabular}
\end{center}

\begin{figure}[t]
    \centering
    \includegraphics[width=0.45\textwidth]  {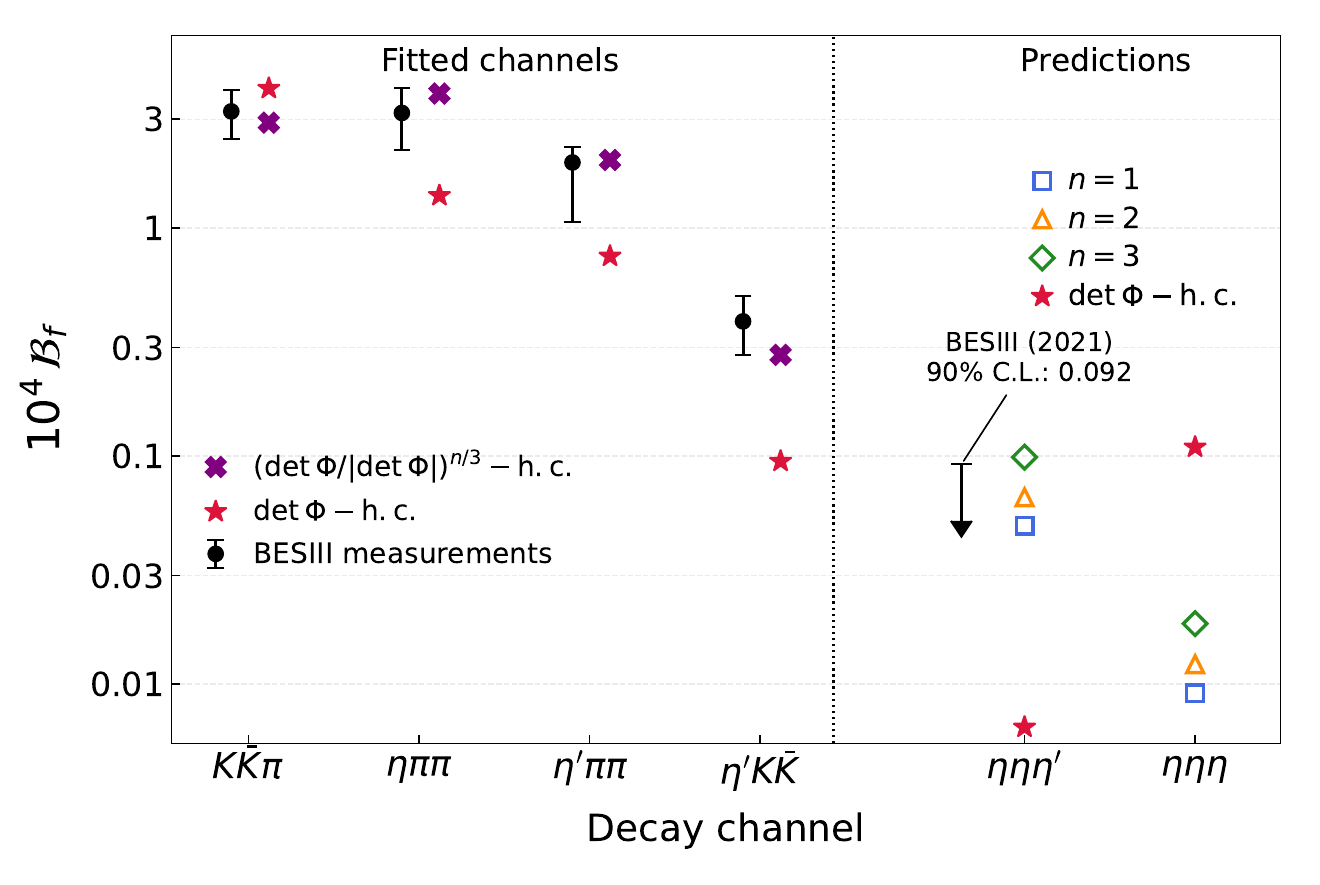}
    \caption{ BESIII data for $X(2370)$ radiative production compared with theory (Eq. (\ref{eq:Pmesons})). On the left, fit results are identical for each harmonic $n=1,2,3$. The $\eta\eta\eta'$ point and $\eta\eta\eta$ are predictions for $n=1,2,3$;  the $\eta\eta\eta'$ BESIII 2021 upper limit at 90\% C.L \cite{BESIII:2020ymv}. For comparison, the results of the plain determinant of Ref. \cite{Eshraim:2012jv} are also presented.}
    \label{fig:X(2370)}
\end{figure}

Within the direct-three-pseudoscalar approximation, we fit one overall
normalization to the measured products
$\mathcal{B}(J/\psi\to\gamma X)\mathcal{B}(X\to f)$
\cite{BESIII:2026rzt,BESIII:2026mvn}.
We evaluate the three-body decays following the standard procedure; see supplemental material.  
The results, in units of $10^{-4}$, are given in the table
above and in Fig. \ref{fig:X(2370)}.
Remarkably, we need a single
parameter to describe very well all four data points, as the $\chi^2/\text{d.o.f}=1.82/3$ shows, supporting a pseudoscalar-glueball interpretation of
$X(2370)$. 

In contrast, using the same inputs and fitting procedure, the
individual frozen-denominator interactions
$iP[(\det\Phi)^{n/N_c}-(\det\Phi^\dagger)^{n/N_c}]$ yield
$\chi^2/\mathrm{d.o.f.}=10.1/3$, $18.7/3$, and $13.5/3$
for $n=1,2,3$, respectively. The last case corresponds to the
conventional determinant interaction studied in
Ref.~\cite{Eshraim:2012jv}, see Fig. \ref{fig:X(2370)}.
This comparison stresses the importance of retaining the field
dependence of $|\det\Phi|$ when calculating decay vertices from the normalized determinant phase.


The channels $\eta\eta\eta$ and $\eta\eta\eta'$ receive both
single-trace and triple-trace contributions and therefore test the
harmonic content. With the normalization fixed by the four-channel
fit, the same direct-decay approximation gives the following product
branching fractions
$\mathcal{B}(J/\psi\to\gamma X)\mathcal{B}(X\to f)$
(in units of $10^{-6}$), without additional fit parameters:
\begin{center}
\begin{tabular}{c|ccc}
channel & $n=1$ & $n=2$ & $n=3$\\
\hline
$\eta\eta\eta$  & $0.91$ & $1.22$ & $1.84$\\
$\eta\eta\eta'$ & $4.95$ & $6.60$ & $9.90$
\end{tabular}
\end{center}
Each column corresponds to a single normalized-phase harmonic.
For $\eta\eta\eta'$, BESIII reports an upper limit of
$9.2\times10^{-6}$ at $90\%$ confidence level
\cite{BESIII:2020ymv}. The $n=1,2$ predictions lie below this
limit, while the $n=3$ central value is slightly above it.
Thus,we stress the importance of measuring the $\eta \eta \eta$ and $\eta \eta \eta'$ decay rates.
A future complete analysis should also include glueball--quarkonium
mixing, scalar contributions
$P\to S\mathcal{P}\to\mathcal{P}\mathcal{P}\mathcal{P}$,
and final-state interactions.

In conclusion, the $\theta$ dependence of the QCD vacuum naturally
leads to a tower of fractional determinant phases, providing a unified
description of anomalous meson and glueball interactions. These
interactions open several directions for future studies, including
hadron phenomenology, the QCD phase diagram, CP violation, and axion
physics.

\vfill\eject
\section*{Acknowledgements}
SJ acknowledges support by the U.S.\ Department of Energy under grants No.\ DE-SC0019232 and No.\ DE-SC0023598.
GyK is supported by the Polish National Science Centre (NCN) under OPUS Grant No. 2022/45/B/ST2/01527. 
PK acknowledges support from the Hungarian National Research, Development and Innovation Fund under Project number K 138277.
RDP is supported by the U.S.
Department of Energy under contract DE-SC0012704, and by the Alexander
von Humboldt Foundation. FR acknowledges support by the Deutsche Forschungsgemeinschaft (DFG, German Research Foundation) through the CRC-TR 211 ``Strong-interaction matter under extreme conditions'' project number 315477589 TRR 211.

\clearpage
\onecolumngrid
\section*{Supplemental Material}

\subsection*{Field definitions}

We summarize here the calculation of the direct three-pseudoscalar decays presented in the main text. The mesonic field is written as
\begin{equation}
 \Phi=S+i\mathcal{P},
\end{equation}
with the scalar and pseudoscalar fields 
\begin{equation}
 S=\frac{1}{\sqrt2}
 \begin{pmatrix}
  (f_0^N+a_0^0)/\sqrt2 & a_0^+ & {K_0^\star}^+\\
  a_0^- & (f_0^N-a_0^0)/\sqrt2 & {K_0^\star}^0\\
  {K_0^\star}^- & \bar{K}_0^\star{}^0 & f_0^S
 \end{pmatrix},
 \qquad
 \mathcal{P}=\frac{1}{\sqrt2}
 \begin{pmatrix}
  (\eta_N+\pi^0)/\sqrt2 & \pi^+ & K^+\\
  \pi^- & (\eta_N-\pi^0)/\sqrt2 & K^0\\
  K^- & \bar K^0 & \eta_S
 \end{pmatrix}.
 \label{eq:Pmatrix}
\end{equation}
For our purpose, it is sufficient to set the scalars to the vacuum expectation value  in the isospin-symmetric limit
\begin{equation}
\Phi_0=\frac12\,\mathrm{diag}\!\left(f_\pi,f_\pi,2f_K-f_\pi\right)
 \equiv \frac12\,\mathrm{diag}(f_\pi,f_\pi,F),
 \qquad F\equiv 2f_K-f_\pi,
 \label{eq:Phi0}
\end{equation}
and keep only the fluctuating pseudoscalar fields, hence
\begin{equation}
 \Phi\to\Phi_0+i\mathcal{P}=\Phi_0(\mathbf{1}+A),\qquad A\equiv \Phi_0^{-1}\mathcal{P}.
 \label{eq:Adef}
\end{equation}
In the flavor-symmetric limit $\Phi_0=\frac{f_{\pi}}{2}\mathbf{1}=v\mathbf{1}$.

Next, $\det\Phi$ contains $\det(\mathbf{1}+A)$.
Using Eqs.~\eqref{eq:Phi0} and \eqref{eq:Pmatrix}, the $A$ matrix has the explicit form
\begin{equation}
 A= 
 \begin{pmatrix}
  \dfrac{\eta_N+\pi^0}{f_\pi} & \dfrac{\sqrt2\,\pi^+}{f_\pi} & \dfrac{\sqrt2\,K^+}{f_\pi}\\[2mm]
  \dfrac{\sqrt2\,\pi^-}{f_\pi} & \dfrac{\eta_N-\pi^0}{f_\pi} & \dfrac{\sqrt2\,K^0}{f_\pi}\\[2mm]
  \dfrac{\sqrt2\,K^-}{F} & \dfrac{\sqrt2\,\bar K^0}{F} & \dfrac{\sqrt2\,\eta_S}{F}
 \end{pmatrix},
 \qquad
 \tr A=\frac{2\eta_N}{f_\pi}+\frac{\sqrt2\,\eta_S}{F}.
 \label{eq:Amatrix}
\end{equation}
The physical isoscalar-pseudoscalar states are obtained from
\begin{align}
 \eta = c\,\eta_N+s\,\eta_S,\qquad
 \eta'=-s\,\eta_N+c\,\eta_S,
 \label{eq:etamix}
\end{align}
with $c \equiv\cos\varphi$, $s \equiv\sin\varphi$ and $\varphi=-44.6^\circ$ \cite{Eshraim:2012jv,Amelino-Camelia:2010cem}. Equivalently, the nonstrange and strange fields can be obtained by
\begin{equation}
 \eta_N=c\, \eta-s\, \eta',\qquad
 \eta_S=s\, \eta+c\, \eta'.
\end{equation}

\subsection*{Numerical inputs}

For the numerical evaluation, we use the isospin-symmetric masses
\begin{equation}
 M_X=2.359~{\rm GeV},\qquad
 m_\pi=0.1380~{\rm GeV},\qquad
 m_K=0.4956~{\rm GeV},
\end{equation}
\begin{equation}
 m_\eta=0.5479~{\rm GeV},\qquad
 m_{\eta'}=0.9578~{\rm GeV},
 \label{eq:nummasses}
\end{equation}
where the light-meson masses are based on the current PDG values~\cite{ParticleDataGroup:2026lqr} and the $X(2370)$ mass is the central value of the updated BESIII determination~\cite{BESIII:2026rzt}. We also use the decay constants
\begin{equation}
 f_\pi =0.0922~{\rm GeV},\qquad
 f_K=0.1100~{\rm GeV},\qquad
 F=2f_K-f_\pi=0.1278~{\rm GeV},
 \label{eq:numdecayconstants}
\end{equation}


\subsection*{Frozen-modulus limit and the determinant interaction}

As a limit, first we freeze the determinant modulus in Eq. (\ref{eq:Pmesons}) of the main text at its vacuum value. With
\begin{equation}
 \alpha_n\equiv\frac{n}{N_c},
\end{equation}
the interaction can be written, up to an overall normalization absorbed into the coupling, as
\begin{equation}
 \mathcal L_n^{\rm fr}=i g_n^{\rm fr}P
 \left\{\left[\det(\mathbf{1}+iA)\right]^{\alpha_n}
 -\left[\det(\mathbf{1}-iA)\right]^{\alpha_n}\right\}.
 \label{eq:frozen}
\end{equation}
At cubic order,
\begin{equation}
\left.\Im\left[\det(1+iA)\right]^{\alpha_n}\right|_{\mathcal{P}^3}
 =-\frac{\alpha_n}{3}\tr A^3
 +\frac{\alpha_n^2}{2}\tr A^2\,\tr A
 -\frac{\alpha_n^3}{6}(\tr A)^3.
 \label{eq:frozenexp}
\end{equation}
Factoring the common coefficient $-\alpha_n/3$ into a redefined coupling gives the reduced cubic operators
\begin{align}
 \mathcal O^{\rm fr}_{1}&=\tr A^3-\frac12\tr A^2\tr A+\frac1{18}(\tr A)^3,
 \label{eq:frozen1}\\
 \mathcal O^{\rm fr}_{2}&=\tr A^3-\tr A^2\tr A+\frac2{9}(\tr A)^3,\\
 \mathcal O^{\rm fr}_{3}&=\tr A^3-\frac32\tr A^2\tr A+\frac12(\tr A)^3.
 \label{eq:frozen123}
\end{align}
If one further imposes the flavor-symmetric vacuum $\Phi_0=v\mathbf 1$, then $A=\mathcal{P}/v$.  For $n=N_c=3$,
\begin{equation}
 \mathcal O^{\rm fr}_{3}
 =\tr \mathcal{P}^3-\frac32\tr \mathcal{P}^2\tr \mathcal{P}+\frac12(\tr \mathcal{P})^3
 =3\det\mathcal{P},
 \label{eq:newton}
\end{equation}
Thus, in this frozen-modulus limit, $n=3$ has the same cubic flavor structure as the conventional determinant interaction
\begin{equation}
 \mathcal L_{\det}=i c_{\text{det}} P\left(\det\Phi-\det\Phi^\dagger\right).
 \label{eq:detlag}
\end{equation}
A separate one-parameter fit of the $n=1,2,3$ frozen-modulus patterns in Eqs.~\eqref{eq:frozen1}--\eqref{eq:frozen123} to the four measured BESIII channels gives, respectively,
\begin{equation}
 \frac{\chi^2}{\mathrm{d.o.f.}}\simeq\frac{10.1}{3},\qquad
 \frac{18.7}{3},\qquad
 \frac{13.5}{3}.
 \label{eq:frozenchi}
\end{equation}
Writing the four-channel reduced pattern as
\begin{equation}
  r_n \equiv
 \left(1,\,
 \frac{\Gamma_{\eta\pi\pi}}{\Gamma_{K\bar K\pi}},\,
 \frac{\Gamma_{\eta'\pi\pi}}{\Gamma_{K\bar K\pi}},\,
 \frac{\Gamma_{\eta'K\bar K}}{\Gamma_{K\bar K\pi}}\right),
 \label{eq:rndef}
\end{equation}
the frozen-modulus results are
\begin{align}
 \bm r^{\rm fr}_1&=(1,\,0.9335,\,0.1675,\,0.02428),
 &\mathcal N^{\rm fr}_1&=3.644,\\
 \bm r^{\rm fr}_2&=(1,\,0.5996,\,9.13\times10^{-5},\,3.09\times10^{-6}),
 &\mathcal N^{\rm fr}_2&=3.617,\\
 \bm r^{\rm fr}_3&=(1,\,0.33935,\,0.18354,\,0.02320),
 &\mathcal N^{\rm fr}_3&=4.114.
 \label{eq:frozenrN}
\end{align}
Here $\mathcal N_n$ is the fitted normalization in the product-branching-fraction units defined below.

\subsection*{B. Full phase-normalized interaction}

We next retain the field dependence of the determinant modulus,
\begin{equation}
 \mathcal L_n=i b_n\PG\left[
 \left(\frac{\det\Phi}{|\det\Phi|}\right)^{n/N_c}
 -\left(\frac{\det\Phi^\dagger}{|\det\Phi|}\right)^{n/N_c}
 \right].
 \label{eq:fulllag}
\end{equation}
Since
\begin{equation}
 \frac{\det\Phi}{|\det\Phi|}=e^{i\vartheta},\qquad
 \vartheta=\Im\tr\ln(1+iA)=\tr A-\frac13\tr A^3+\mathcal O(A^5),
\end{equation}
we obtain
\begin{equation}
 \left.\Im\left[
 \left(\frac{\det\Phi}{|\det\Phi|}\right)^{n/N_c}
 \right]\right|_{P^3}
 =-\frac{n}{3N_c}\tr A^3
 -\frac{n^3}{6N_c^3}(\tr A)^3.
 \label{eq:fullcubic}
\end{equation}
The mixed structure $\tr A^2\tr A$ cancels exactly against the expansion of the inverse modulus. After absorbing the common factor $-n/(3N_c)$ into the coupling, the three explicit reduced interactions for $N_c=3$ are
\begin{align}
 \mathcal L_{1}^{(3)}=g_1\PG\left[\tr A^3+\frac1{18}(\tr A)^3\right],\\[1mm]
\mathcal L_{2}^{(3)}=g_2\PG\left[\tr A^3+\frac2{9}(\tr A)^3\right],\\[1mm]
 \mathcal L_{3}^{(3)}=g_3\PG\left[\tr A^3+\frac12(\tr A)^3\right].
 \label{eq:full123}
\end{align}
For reference, the trace coefficients are summarized in Table~\ref{tab:tracecoeff}.
\begin{table}[h]
\centering
\caption{Reduced coefficients multiplying the three cubic trace structures after the common overall coupling has been factored out.}
\label{tab:tracecoeff}
\begin{tabular}{c|ccc}
\toprule
interaction & $\tr A^3$ & $\tr A^2\tr A$ & $(\tr A)^3$\\
\midrule
full $n=1$   & $1$ & $0$ & $1/18$\\
full $n=2$   & $1$ & $0$ & $2/9$\\
full $n=3$   & $1$ & $0$ & $1/2$\\
\midrule
frozen $n=1$ & $1$ & $-1/2$ & $1/18$\\
frozen $n=2$ & $1$ & $-1$   & $2/9$\\
frozen $n=3$ & $1$ & $-3/2$ & $1/2$\\
\bottomrule
\end{tabular}
\end{table}

\subsection*{C. Flavor coefficients entering the three-body widths}

For the full interaction, we define
\begin{equation}
 q_n\equiv\frac12\alpha_n^2
 =\left\{\frac1{18},\frac2{9},\frac12\right\},\qquad n=1,2,3,
\end{equation}
so that the reduced flavor operator is
\begin{equation}
 \mathcal O_n=\tr A^3+q_n(\tr A)^3.
\end{equation}
The coefficient of a monomial in $\mathcal O_n$ is denoted $C_f^{(n)}$. The entries below follow directly from the explicit matrix $A$ in Eq.~\eqref{eq:Amatrix}.  Table~\ref{tab:physicalcoeff} gives the coefficients used in the decay calculation. For rows containing two identical fields, the listed neutral coefficient is the Feynman-rule coefficient after differentiating with respect to the two identical fields; the identical-particle factor is treated consistently in the phase-space formula.

\begin{table}[h]
\centering
\caption{Flavor coefficients for representative charge channels. The first four inclusive channels receive no contribution from $(\tr A)^3$.}
\label{tab:physicalcoeff}
\renewcommand{\arraystretch}{1.25}
\begin{tabular}{cc}
\hline
final state & $C^{(n)}$\\
\hline
$K^+K^-\pi^0$
& $\displaystyle \frac{6}{f_\pi^2F}$\\

$K^0\bar K^0\pi^0$
& $\displaystyle -\frac{6}{f_\pi^2F}$\\

$K^+\bar K^0\pi^-$, $K^-K^0\pi^+$
& $\displaystyle \frac{6\sqrt2}{f_\pi^2F}$\\

$\eta\pi^+\pi^-$, $\eta\pi^0\pi^0$
& $\displaystyle \frac{12c}{f_\pi^3}$\\

$\eta'\pi^+\pi^-$, $\eta'\pi^0\pi^0$
& $\displaystyle -\frac{12s}{f_\pi^3}$\\

$\eta'K^+K^-$, $\eta'K^0\bar K^0$
& $\displaystyle \frac{6(\sqrt2\,cf_\pi-Fs)}{f_\pi^2F^2}$\\\hline
\end{tabular}
\end{table}

For the two all-isoscalar channels both traces contribute. Using
\begin{equation}
 \left.\tr A\right|_{\eta,\eta'}=
 \left(\frac{2c}{f_\pi}+\frac{\sqrt2\,s}{F}\right)\eta
 +\left(-\frac{2s}{f_\pi}+\frac{\sqrt2\,c}{F}\right)\eta',
\end{equation}
the monomial coefficients before the identical-field factorials are
\begin{align}
 C_{\eta^3}^{(n)}&=
 \frac{2c^3}{f_\pi^3}+\frac{2\sqrt2\,s^3}{F^3}
 +q_n\left(\frac{2c}{f_\pi}+\frac{\sqrt2\,s}{F}\right)^3,
 \label{eq:ceta3}\\
 C_{\eta^2\eta'}^{(n)}&=
 -\frac{6c^2s}{f_\pi^3}+\frac{6\sqrt2\,s^2c}{F^3}
 +3q_n\left(\frac{2c}{f_\pi}+\frac{\sqrt2\,s}{F}\right)^2
 \left(-\frac{2s}{f_\pi}+\frac{\sqrt2\,c}{F}\right).
 \label{eq:ceta2etap}
\end{align}
The corresponding Feynman-rule coefficients contain factors $3!$ for $\eta^3$ and $2!$ for $\eta^2\eta'$. These are precisely the channels in which the $n$ dependence survives after the overall normalization has been fixed.

\subsection*{D. Three-body decay formula}

We evaluate the direct three-body widths using the standard phase-space formula reported in the kinematics of the PDG \cite{ParticleDataGroup:2026lqr} and following the notation of Ref~\cite{Eshraim:2012jv}. Below, $X$ refers to the glueball candidate $X(2370)$
\begin{equation}
 \Gamma_{X\to P_1P_2P_3}
 =\frac{s_{123}}{32(2\pi)^3M_G^3}
 \int_{(m_1+m_2)^2}^{(M_X-m_3)^2}dm_{12}^2
 \int_{(m_{23}^2)_{\min}}^{(m_{23}^2)_{\max}}dm_{23}^2\,
 \left|-i\mathcal M_{X\to P_1P_2P_3}\right|^2,
 \label{eq:width3}
\end{equation}
where $s_{123}=1,2,6$ for three different particles, two identical particles, and three identical particles. The Dalitz boundaries are
\begin{align}
 (m_{23}^2)_{\min,\max}
 &= (E_2^*+E_3^*)^2
 -\left(\sqrt{E_2^{*2}-m_2^2}\,\pm\sqrt{E_3^{*2}-m_3^2}\right)^2,\\
 E_2^*&=\frac{m_{12}^2-m_1^2+m_2^2}{2m_{12}},\qquad
 E_3^*=\frac{M_X^2-m_{12}^2-m_3^2}{2m_{12}}.
\end{align}
The local interactions considered here give momentum-independent tree-level amplitudes, so each width is determined by the flavor coefficient, charge multiplicity, identical-particle bookkeeping, and the physical three-body phase space.

\subsection*{E. Fit to the experiment}

The experimentally quoted quantities are not, by themselves, the absolute branching fractions $\BR[X(2370)\to f]$. BESIII measures product branching fractions
\begin{equation}
 P_f\equiv 10^4\,\BR[J/\psi\to\gamma X(2370)]\,\BR[X(2370)\to f].
 \label{eq:productBF}
\end{equation}
The 2026 BESIII measurements and summary provide the four channels used in the fit, while the 2021 BESIII analysis gives the upper limit for $\eta\eta\eta'$~\cite{BESIII:2020ymv,BESIII:2026mvn}. 

The Lagrangian determines the relative partial-width pattern
\begin{equation}
 r_f^{(n)}\equiv\frac{\Gamma_f^{(n)}}{\Gamma_{K\bar K\pi}^{(n)}}.
\end{equation}
We fit one common normalization $\mathcal N_n$ for each model,
\begin{equation}
 P_f^{\rm th}=\mathcal N_n\,r_f^{(n)},
 \label{eq:fitform}
\end{equation}
to the four measured product branching fractions. For the full phase-normalized interaction the first four channels depend only on $\tr A^3$, so the common factor $n/N_c$ is absorbed into $\mathcal N_n$ and all three values $n=1,2,3$ give the same fitted pattern.  The fit used in the main text gives
\begin{equation}
 \frac{\chi^2}{\mathrm{d.o.f.}}=\frac{1.82}{3},
\end{equation}
The common reduced pattern and fitted normalization are
\begin{equation}
  \bm r=(1,\,1.34093,\,0.68587,\,0.09603),
 \qquad
 \mathcal N=2.90247.
 \label{eq:fullrN}
\end{equation}
This gives the fitted product branching fractions shown below,
\begin{center}
\begin{tabular}{c|c|c}
\toprule
channel & BESIII & theory\\
\midrule
$K\bar K\pi$ & $3.25^{+0.77}_{-0.79}$ & $2.902$\\
$\eta\pi\pi$ & $3.20^{+0.91}_{-1.00}$ & $3.892$\\
$\eta'\pi\pi$ & $1.94^{+0.33}_{-0.88}$ & $1.991$\\
$\eta'K\bar K$ & $0.39\pm0.11$ & $0.279$\\
\bottomrule
\end{tabular}
\end{center}
all in the units of Eq.~\eqref{eq:productBF}.

Once $\mathcal N_n$ has been fixed by these four modes, the all-isoscalar channels are genuine predictions. We obtain approximately
\begin{center}
\begin{tabular}{c|ccc}
\toprule
channel & $n=1$ & $n=2$ & $n=3$\\
\midrule
$\eta\eta\eta$ & $0.0091$ & $0.0122$ & $0.0184$\\
$\eta\eta\eta'$ & $0.0495$ & $0.0661$ & $0.0990$\\
\bottomrule
\end{tabular}
\end{center}
in the same product-branching-fraction units. BESIII reported
\begin{equation}
 10^4\,\BR[J/\psi\to\gamma X(2370)]\,\BR[X(2370)\to\eta\eta\eta']<0.092
 \qquad (90\%~\mathrm{C.L.})
\end{equation}
in 2021~\cite{BESIII:2020ymv}.

\twocolumngrid
\bibliography{bib}
\end{document}